\documentclass[12pt]{article}

\usepackage{graphicx}
\usepackage{subfigure}
\usepackage[hang]{caption2}

\def\be{\begin{equation}}
\def\ee{\end{equation}}
\def\ba{\begin{eqnarray}}
\def\ea{\end{eqnarray}}

\def\ltsim{\ {\raise-3pt\hbox{$\sim$}}\!\!\!\!\!{\raise2pt\hbox{$<$}}\ }
\def\gtsim{\ {\raise-3pt\hbox{$\sim$}}\!\!\!\!\!{\raise2pt\hbox{$>$}}\ }

\begin{document}

\begin{titlepage}

\vskip -0.2in

\rightline{OSU-HEP-98-1}
\rightline{UTEXAS-HEP-98-1}
\rightline{DOE-ER-40757-108}

\vskip 0.5in

\begin{center}

{\large \bf Signatures for the Charged Higgs Decay of the Top Quark at the
Tevatron}
\vskip 0.3in
{\bf Duane A. Dicus}\footnote{e-mail address: phbd057@utxvms.cc.utexas.edu}\\

\vskip 0.2in

{\em Center for Particle Physics,\\ Department of Physics,\\ 
University of Texas at Austin,\\
Austin, TX 78712}\\

\vskip 0.2in

{\bf David J. Muller}\footnote{e-mail address: qgd@okstate.edu} and
{\bf Satyanarayan Nandi}\footnote{e-mail address: shaown@vms.ucc.okstate.edu}\\

\vskip 0.2in

{\em Department of Physics,\\ Oklahoma State University,\\
Stillwater, OK 74078}\\

\vskip 0.5in

{\bf ABSTRACT}

\end{center}

\begin{quotation}

We investigate the effect that a charged Higgs decay channel for the top
quark has on the signature for top quark pair production at the 
Tevatron. The branching ratios for the various  multijet and multilepton
final states arising from $t \overline{t}$ production
are obtained
for a wide range of values for $\tan \beta$ and the charged Higgs boson 
mass. In addition, the effect of the charged Higgs decay of the top
quark on the 2 b-tagged jets and 1 lepton channel is considered.

\end{quotation}

\end{titlepage}

\section{Introduction}

The discovery of the top quark by the CDF and D0 collaborations potentially
opens a window for the discovery of new physics beyond the standard
model. With the top quark's discovery, all the particles of the standard model 
have been found with the exception of the Higgs boson. Since the top quark's 
mass is so
much larger than that of the other known fermions, the physics of the top
should provide a useful window into the electroweak symmetry breaking 
mechanism. It can also provide information on new TeV scale physics such as 
low energy supersymmetry, especially if some of the 
new particles are less massive
than the top quark and can thereby be directly produced.

One popular scenario for new physics involves extended Higgs sectors. This is
particularly true in supersymmetric theories where more than one Higgs doublet
is required for anomaly cancellation and to give masses to both up-type and
down-type quarks. 
The minimal supersymmetric standard model (MSSM) takes the minimal Higgs
structure of two doublets.
Of the eight degrees of freedom in this model, three are absorbed to
give mass to the W and Z bosons which leaves three neutral Higgs bosons and
a charged Higgs boson pair. If the charged Higgs boson is light enough, 
then it
is possible that the top quark has a nonstandard decay mode into these
bosons. If this is the case, then top quark events observed at colliders
such as the Tevatron would differ from the standard model (SM) expectation. 
Thus, the
charged Higgs boson will either be detected through top quark decay or some
bound will be placed on the mass of the charged Higgs boson through its
nondetection.

Numerous limits have already been placed on the values that the mass of
the charged Higgs boson and $\tan \beta$ can take. One such limit can be
obtained from the following relation which
holds at tree level
\be
M_{H^\pm}^2 = M_{A^\circ}^2 + M_W^2 \ 
\ee
where $M_{H^\pm}$ is the mass of the charged Higgs boson and $M_{A^\circ}$
is the mass of the pseudoscalar Higgs.
The current ALEPH 95\% C.L. limit of $M_{A^\circ} > 62.5$\,GeV 
for $\tan \beta > 1$ [\ref{cowan}] 
then implies that
$M_{H^\pm} > 100$\,GeV. 
For such values of $\tan \beta$, the one-loop corrections tend to shift
the charged Higgs mass down from its tree-level value by less than 
10\,GeV. The size of the correction decreases with increasing $\tan \beta$
[\ref{diaz}].
Moreover, limits from the nonobservance of direct pair
production of charged Higgs bosons at LEP (including LEP2) 
set lower bounds on the charged Higgs mass. At the 95\% confidence level, 
the DELPHI collaboration sets a lower bound of 54.4\,GeV, ALEPH
sets a lower bound of 52\, GeV, and OPAL sets a lower
bound of 52.0\,GeV [\ref{direct}].

Bounds on the values for $M_{H^\pm}$ and $\tan \beta$ have also 
been obtained by considering the charged Higgs contribution to inclusive 
semi-tauonic $B$-decays [\ref{grossman}]. Recently, the supersymmetric
short-distance QCD corrections have been incorporated into the analysis.
Using the current bounds on the sparticle masses, 
the bound 
\be
\tan \beta  \ltsim 0.43 (M_{H^\pm}/{\rm GeV}) 
\ee
at the
$2 \sigma$ level for $\mu < 0$ is obtained (the $\mu$ term in the 
superpotential is taken to be $-\mu H_1 H_2$). For $\mu > 0$ these decays 
could yield no bound at all [\ref{sola}].

The CDF collaboration at the Tevatron has searched for charged Higgs decays
of the top quark [\ref{CDF}]. Recently they have searched for evidence of
such decays by considering hadronic decays of the tau lepton since the 
charged Higgs decays primarily to the tau for
$\tan \beta > 4$. Seven events meet their cuts with an expected background
of $7.4 \pm 2.0$ events. A region in the $\tan \beta - M_{H^\pm}$ plane
is thereby excluded. In particular, charged Higgs bosons with $M_{H^\pm}
< 147(158)$\,GeV are excluded in the large $\tan \beta$ limit 
($\tan \beta > 100$) for a top 
quark mass of 175 GeV and top
production cross section $\sigma_{t \overline{t}} = 5.0(7.5)$\,pb.
Moreover, to maintain consistency with the then observed top quark 
cross section of
$\sigma_\circ = 6.8^{+3.6}_{-2.4}$\,pb, $\sigma_{t \overline{t}}$ must
increase at higher $\tan \beta$ to compensate for the lower branching 
ratio into the SM mode Br($t \overline{t} \rightarrow Wb W\overline{b}$). 
This excludes more of the parameter space [\ref{CDF}].

Similarly, Guchait and Roy have used Tevatron top quark data in the lepton
plus $\tau$ channel to obtain a significant limit on the $H^{\pm}$ mass in the
large $\tan \beta$ region [\ref{roy}]. They consider the lepton plus multijet 
channel looking for departures from the SM prediction due 
to the charged Higgs' preferential coupling to the tau lepton. They thereby
obtain an exclusion area in the $\tan \beta - M_{H^\pm}$ plane. 
Quantitatively, they obtain a mass limit of 100\,GeV for $\tan \beta \geq
40$ increasing to 120\,GeV at $\tan \beta \geq 50$. Essentially the same
analysis was performed by Guasch and Sol\`{a} but with the MSSM quantum
corrections included [\ref{guasch}]. They demonstrated that these 
corrections have a
substantial impact on the allowed parameter space. In particular, for 
$\mu > 0$, these corrections decrease the cross section for the
$\tau$ signal and a light charged Higgs mass ($\sim 100$\,GeV) could hold
for essentially any (perturbative) value of $\tan \beta$.

In this paper, we investigate the signature for top quark production at
hadronic colliders for the case where the charged Higgs boson is light 
enough for the top quark to decay into it. We determine the branching
ratios for the decays into various numbers of jets and leptons as a 
function of $\tan \beta$ and the charged Higgs mass.

\section{Theory}

The MSSM contains two Higgs doublets. One Higgs doublet couples to the up-type 
quarks and the neutrinos, while the other couples to the down-type quarks and 
the charged leptons. Three of the eight degrees of freedom are absorbed to 
give mass to the $W$ and $Z$ bosons. This leaves five physical Higgs bosons:
three neutral ($h^\circ$, $H^\circ$, and $A^\circ$) and a charged pair
($H^\pm$).
If the charged Higgs is lighter than the top quark, then it is
possible that the top quark decays in part through the charged Higgs. The
allowed decays for the top quark in the MSSM are then $t \rightarrow b W^+$
and $t \rightarrow b H^+$; there are no other decay modes ignoring 
intergenerational mixing. The interactions of
the charged Higgs bosons with quarks are represented by the Lagrangian:
\be
{\cal L} = \frac{g}{2 \sqrt{2} M_W} H^+ \left[\cot \beta \overline{U} M_U D_L
 + \tan \beta \overline{U} M_D D_R \right] + h.c. 
\ee
where U represents the three generations of up-type quarks and D represents 
the three generations of down-type quarks.
$M_U$ and $M_D$ are diagonal up and down quark mass matrices. We have
set the CKM matrix to the identity matrix since we are neglecting the
small intergenerational mixings for our analysis.
$\tan \beta \equiv v_2/v_1$ where 
$v_1$ is the vacuum expectation value (vev) for the Higgs doublet that
couples to the down-type quarks and $v_2$ is the vev for the Higgs doublet
that couples to up-type quarks.
The widths for the top quark's decays are then
\ba
\nonumber \Gamma (t \rightarrow b W) & = & \frac{g^2}{64 \pi M_W ^2 M_t} 
\lambda^{\frac{1}{2}}(1,\frac{M_b^2}{M_t^2},\frac{M_W^2}{M_t^2}) \times \\
& & \left[M_W^2(M_t^2 + M_b^2) + (M_t^2 - M_b^2)^2 - 2M_W^4 \right] \\
\nonumber \Gamma (t \rightarrow b H) & = & \frac{g^2}{64 \pi M_W^2 M_t}
\lambda^{\frac{1}{2}}(1,\frac{M_b^2}{M_t^2},\frac{M_H^2}{M_t^2}) \times \\
& &\left[ (M_t^2 \cot^2 \beta + M_b^2 \tan^2 \beta) 
 (M_t^2 + M_b^2 - M_H^2) + 4 M_t^2 M_b^2 \right]
\ea
where $\lambda(x,y,z) \equiv x^2 + y^2 + z^2 - 2xy - 2xz - 2yz$. 
The
branching fraction for $t \rightarrow b H^+$ is large (\,$>$\,10\%\,) for
$\tan \beta \leq 1$ and $\tan \beta > \frac{M_t}{M_b}$.

The charged Higgs in turn decays into the standard fermions. Its coupling 
to the fermions increases with their mass, so the primary decay modes to
consider for the charged Higgs are $H^+ \rightarrow c \overline{s}$ and
$H^+ \rightarrow \nu_\tau \overline{\tau}$. The widths for these decays are
\ba
\nonumber \Gamma (H^+ \rightarrow c \overline{s}) & = & 
\frac{3 g^2}{32 \pi M_H M_W^2} \lambda^{1/2}(1,\frac{M_c^2}{M_H^2},
\frac{M_s^2}{M_H^2}) \times \\ 
& & \left[(M_s^2 \tan^2 \beta
+ M_c^2 \cot^2 \beta)(M_H^2 - M_c^2 - M_s^2) - 4 M_c^2 M_s^2 \right] \\
\Gamma (H^+ \rightarrow \nu_\tau \overline{\tau}) & = &
\frac{g^2}{32 \pi M_H^3 M_W^2}(M_H^2 - M_\tau^2)^2 M_\tau^2 \tan^2 \beta \ .
\ea
For small $\tan \beta$, the decay $H^+ \rightarrow c \overline{s}$ dominates,
while for large $\tan \beta$, the decay $H^+ \rightarrow \nu_\tau
\overline{\tau}$ dominates. The branching ratio for $H^+ \rightarrow
\nu_\tau \overline{\tau}$ is essentially unity for $\tan \beta > 4$.

\section{Analysis and Results}

In this analysis we study the possible Tevatron signatures for charged
Higgs production through top quark decay in 
the context of the MSSM.
The cuts employed are that final state charged leptons (electrons and
muons) must have a $p_T$ greater than 20 GeV and a pseudorapidity, 
$\eta \equiv -\ln (\tan \frac{\theta}{2})$ (where $\theta$ is the polar angle
with respect to the proton beam direction), 
of magnitude less than 1. 
Jets must have an $E_T > 15$\,GeV and $|\eta| < 2$. In addition, 
hadronic final states within a cone size of
$\Delta R \equiv \sqrt{ (\Delta \phi)^2 + (\Delta \eta)^2} = 0.4$ are
merged to a single jet. 
The signature here for the hadronic decay of the $\tau$ lepton is to a
single thin jet and we assume this is always true.
Leptons within this cone radius of a jet are
discounted. Throughout this analysis, the mass of the top quark is taken to
be 175\,GeV in accordance with current CDF and D0 collaboration 
measurements.

There are several possible final states available from top pair production.
With the two decay possiblilities of $t \rightarrow W^+ b$ and 
$t \rightarrow H^+ b$, there can be up to two b-jets from the decays of the
top quarks. For the $W$ decay channel, which is the only channel 
available to the SM, the $W$ bosons can decay to as many as two jets each or
they can each decay leptonically. Thus in the
SM case, one can expect after implementing the cuts any number of jets up
to six and any number of charged leptons up to 2. Introducing the 
possibility of the top quark decaying via the charged Higgs boson 
changes the branching ratios for decay into these various channels.
For $\tan \beta > 4$, the charged Higgs decays primarily into the $\tau$
lepton. With the hadronic decay of the $\tau$ being to a single thin jet,
there should be a depletion in the number of events with large numbers of
jets when
the branching ratio for $t \rightarrow H^+ b$ becomes
appreciable. 
As $\tan \beta$ falls below $\sim 4$, the branching ratio for 
$H^+ \rightarrow \overline{\tau} \nu_\tau$ decreases and the branching ratio
for $H^+ \rightarrow c \overline{s}$ increases correspondingly. The values
for the branching ratios change particularly sharply for $\tan \beta \sim 1$:
for a charged Higgs mass of 110\,GeV,
the branching ratio for $H^+ \rightarrow \overline{\tau} \nu_\tau$ is 
$\sim 0.94$ for $\tan \beta = 3$ and is $\sim 0.31$ for $\tan \beta = 1$.
Thus for small $\tan \beta$, $H^+ \rightarrow c \overline{s}$ becomes the
dominant decay mode of the charged Higgs and we expect a depletion in
events with leptons.

Events that contain leptons are distinctive. This is particularly true for
final states containing two leptons. While production rates for these
dilepton modes is rather small, their distinctive signature allows for a
good separation from background. Thus the two-jets and two-leptons mode, which
has the largest branching ratio of the dilepton modes, could be useful for
charged Higgs detection after a long collider run. Fig.~1 shows a plot of
the branching ratio versus charged Higgs mass for the two-jets dilepton
mode. Each curve represents a different value for $\tan \beta$. As the 
figure shows, the curves for the various values of $\tan \beta$ all lie
below the SM expectation. Thus if the decay $t \rightarrow H^+ b$
is allowed, there will be a depletion of events for this mode which already
had a small branching ratio in the SM case. For $M_{H^\pm}$ around 110 GeV,
the branching ratio can be as low as half the SM expectation.

The 2-jets dilepton mode occurs when the decays of the $W$ or $H^\pm$ 
from each top quark 
leads to an $e$ or a $\mu$ either directly or indirectly. For 
$\tan \beta > 4$, the predominant decay mode of the charged Higgs is to
the $\tau$ lepton. Since the electrons and muons from the subsequent 
tau decays
occur farther along the decay chain than $e$'s and $\mu$'s from direct
$W$ decay, there is less energy available for the electrons and muons from 
the $\tau$ decays than in $W$ decays. Thus the electrons and muons from the
tau decays tend to be softer and less likely to meet the $p_T$
cuts. Thus as $\tan \beta$ increases (and so as BR($t \rightarrow H^+ b$) 
increases), the branching ratio for the two jet and two
lepton decay mode gets smaller. The branching ratios for this mode increase
to the SM value as $M_{H^\pm}$ increases towards 170\,GeV and the phase 
space available for the charged Higgs decay of the top quark goes to zero.

The depletion in two-jets dilepton events is considerable for low
values of $\tan \beta$ as well.
For $\tan \beta = 1$, Fig.~1 shows that the branching ratio for the
two-jets dilepton mode decreases rapidly as the charged Higgs mass
decreases. This is due to the fact that as $\tan \beta$ falls below
approximately 6, the branching ratio for $t \rightarrow H^+ b$
increases. Moreover, as $\tan \beta$ falls below 3, the branching ratio for
$H^+ \rightarrow c \overline{s}$ increases rapidly. 
Thus as $\tan \beta$ decreases,
events with large numbers of leptons become less plentiful. For 
$\tan \beta = 1$ and $M_{H^\pm} = 110$\,GeV, for example, the branching 
ratio for $H^+ \rightarrow c \overline{s}$ is 69\%. Hence the depletion
in dilepton events with two jets relative to the SM case is due here to a
general lack of leptonic decays of the charged Higgs boson. Those $e$'s
and $\mu$'s coming indirectly from the charged Higgs also tend to be soft
as dicussed above and tend to be eliminated by the $p_T$ cuts.

The single lepton decay modes have the advantage that they are produced
at a more substantial rate than the dilepton modes while retaining some
of the distinctiveness that lepton modes offer. The plots of the branching
ratios versus the charged Higgs mass for these modes are shown in Fig.~2.
Fig.~2a shows the plot for the four jets and one lepton case. 
We can see
that the branching ratios for this mode are all below the SM case.
Moreover, the solid curves show that as the value for $\tan \beta$ 
increases above 5, the branching ratio decreases. As
$\tan \beta$ falls below 5, there is likewise a decrease in the branching
ratio for the 4 jets-1 lepton mode.

We begin our analysis of Fig.~2a by considering the case where 
$\tan \beta > 5$. Two of the jets in the 4 jets-1 lepton mode are typically 
b-jets coming from the decay of the top quarks. This leaves two ways in
which we can obtain a total of four jets and one charged lepton. The first
way is for both of the top quarks to decay via the $W$ with one $W$
decaying hadronically and the other leptonically. The other way is for
one of the top quarks to decay via the $W^\pm$ which subsequently decays
hadronically and the other top decaying via the $H^\pm$ which decays
indirectly to an $e$ or $\mu$ through the $\tau$. 
Other possibilities would involve the 
subsequent hadronic decays of the $\tau$ leptons from charged Higgs decays, 
but these can not contribute to this mode
as they lead to only one jet instead of the required two. Since only a 
subset of the possible top decays can give rise to the 4 jets-1 lepton mode,
there will be a decrease in the number of events in this mode. This 
decrease is made more pronounced by the relative softness of the electrons
and muons coming from the $H^\pm$ decay chain which tends to eliminate them
when the $p_T$ cuts are applied. 
As $\tan \beta$ increases beyond 5, the
branching ratio for this mode decreases because of the increase in charged
Higgs production. 
The minimum deviation from the SM occurs for 
$\tan \beta \sim$ 6 -- 7 as this is where the minimum in the branching ratio
for $t \rightarrow H^+ b$ occurs.

We now consider the case for $\tan \beta < 5$. For $\tan \beta  = 3$, $H^\pm$
production increases somewhat from $\tan \beta = 5$, but the branching 
ratio for $H^+ \rightarrow \overline{\tau} \nu_\tau$ remains close to one.
As a result, we get slightly fewer events for this mode compared to the 
$\tan \beta = 5$
case. For $\tan \beta = 1$, on the other hand, there is not only more $H^\pm$
production than in the $\tan \beta = 3$ case (the increase is by a factor
of $\sim 5$), but the branching ratio for $H^+ \rightarrow c \overline{s}$
has increased from 0.06 to 0.69. Thus there is a decrease in  4jets-1 lepton
events due to a general decrease in leptonic events.

The three jets and single lepton mode, whose branching ratio as a function
of $M_{H^\pm}$ is given in Fig.~2b, shows rather different behavior. In the 
large $\tan \beta$ region, the branching ratios for this mode are all above
the SM expectation. These tend to increase with increasing $\tan \beta$.
These features can be qualitatively understood as
follows. Two of the jets are almost always b-jets coming from the decays
of $t$ and $\overline{t}$. The remaining one jet and one lepton must come
from the decays of the $W$ and charged Higgs bosons. For the 
$\tan \beta > 5$ case, the branching ratios before cuts for obtaining one jet 
(indirectly from $\tau$ decay) and one $e$ or $\mu$ (directly or indirectly
from $\tau$ decay) from $WW$, $WH$, and $HH$ are 0.18, 0.19, and 0.22, 
respectively. Thus, charged Higgs production naturally gives rise to
branching ratios for the 3 jets-1 lepton mode that are larger than the SM
case. There are also other contributing factors for this increase. We have 
seen that the branching ratios for the 4 jets-1 lepton case tend to be 
below the SM value. The events that would have had four jets but failed to
meet the isolation cuts for two of the jets could be taken as a three jets and
one lepton event.  Finally, the branching ratios for the 3 jets and 
1 lepton mode 
increase as $\tan \beta$ increases due to the corresponding increase in 
BR($t \rightarrow H^+ b$).

The branching ratio curves in Fig.~2b also show another interesting
feature. For large values of $\tan \beta$, as the charged Higgs mass 
decreases from 170\,GeV, the branching ratios increase. However, this 
increase
flattens out and the the branching ratios start to decrease
around 140\,GeV. This decrease becomes sharper around 120\,GeV through
100\,GeV. This decrease is due to the increase in the branching ratio
for $t \rightarrow H^+ b$ as the charged Higgs mass decreases so
that more of the leptons are soft leptons from charged Higgs decay which
tend to be more easily eliminated by the $p_T$ cut. In any case, the various
factors that try to either increase or decrease the branching ratio
for the 3 jets-1 lepton mode tend to keep the branching ratios roughly
constant. The $\tan \beta = 80$ branching ratio varies by less than 15\%
over the mass range.
 
For the low $\tan \beta$ region, the branching ratios for the 3 jets and
1 lepton mode tend to fall below the SM expectation as shown by the
$\tan \beta = 1$ and 1.25 curves in Fig.~2b. 
This is due to the general decrease
in events containing leptons as $\tan \beta$ falls below 4 and the branching
ratio for $H^+ \rightarrow c \overline{s}$ increases. Indeed, for 
$\tan \beta \simeq 1$ and a charged Higgs mass of 110\,GeV,
$H^+ \rightarrow c \overline{s}$ is the dominant decay mode with a branching
ratio of 69\%. The graph shows a corresponding drop of 15\% in the
branching ratio below the SM value.

Fig.~2c shows the branching ratios for the two jets and one lepton mode.
The values of the branching ratios are larger than the SM case except for
$\tan \beta = 1$ and increase with $\tan \beta$ for $\tan \beta > 5$.
In this mode, two of the jets are almost always the b-jets coming from the
decays of $t$ and $\overline{t}$. 
The main  process generating the 2 jets-1 lepton events is then for both
the $W$ and charged Higgs bosons to decay via the leptonic mode and
then have one of the charged leptons (typically from the charged Higgs)
fail to meet the cuts.
For
$\tan \beta > 5$, the leptonic branching ratios for the $WW$, $WH$, and $HH$
decays are 0.07, 0.18, and 0.12, respectively. Thus, Higgs production
gives rise to branching ratios for this mode that are larger than the SM
expectation. As $\tan \beta$ increases, the branching ratios increase due
to the increasing branching ratio for $t \rightarrow H^+ b$. 
For $\tan \beta = 3$, $H^\pm$ production increases somewhat from
$\tan \beta = 5$, but the branching ratio for 
$H^+ \rightarrow \overline{\tau} \nu_\tau$ is still close to one. As a result,
the branching ratio for the 2 jets and 1 lepton mode is somewhat larger than
for the $\tan \beta = 5$ case. The $\tan \beta = 1$ curve is below the SM
case due to a decrease in leptonic events as $H^+ \rightarrow c \overline{s}$
is the dominant decay mode for the charged Higgs.

Events with purely hadronic final states are less interesting due to the 
fact that they are harder to separate from the background at hadronic
colliders. Nevertheless, these events can still be a source of interesting
information on charged Higgs decays of the top. Fig.~3 shows the branching 
ratios for the purely hadronic modes. Figs.~3a and 3b show the 6 jets and 5 
jets modes, respectively. As they show, the branching ratios tend to fall
below the SM case. This is again due to the general depletion in events
with large numbers of jets as the 
rate for top quark decay to the charged Higgs
is increased. Fig.~3c shows the 4 jets case. Like the 3 jets-1 lepton case,
this case has the branching ratios being roughly constant. Here this is 
due to a decrease in large jet activity being balanced by events
failing to meet the cuts in the 5 jets-0 lepton mode and the 4 jets-1 lepton
modes (an increase in events with charged Higgs bosons means an increase in
softer leptons that will be less likely to meet the $p_T$ cuts). 
Figs.~3d and 3e show the branching ratios for the 3 jets and 2 jets modes.
Here the branching ratios are all above the SM case and increase with
increasing $\tan \beta$ and decreasing $M_{H^\pm}$. 
Thus, a significant enhancement in the dijet production where both jets
are high $p_T$ b-tagged jets could be an interesting signal for charged
Higgs production.
For $\tan \beta > 4$, this is yet again due to the fact that increasing the
probability for $t \rightarrow H^+ b$ increases the number of events with
smaller numbers of jets. For $\tan \beta < 4$, the increase is due to a
general depletion of events with charged leptons.

In order to gain a sense of the size of the branching ratios of the various
decay modes relative to each other, histograms of the branching ratios
for a few representative cases are shown in Fig.~4. The SM expectation is 
given in Fig.~4a. Figs.~4b and 4c show the case where $M_{H^\pm} = 110$\,GeV
and $\tan \beta = 3$ and 50, respectively. The $\tan \beta = 3$ histogram
is quite similar to the SM case owing to the fact that the top quark decays
primarily via the $W$ boson for this value of $\tan \beta$. We can see some
depletion in the 6 jets-0 lepton bin as a result of the depletion in 
large jet
activity associated with charged Higgs production where the charged Higgs
decays primarily via the tau lepton. This effect is much more pronounced in 
the $\tan \beta = 50$ case as shown in Fig.~4c. Here the branching ratio for
top decay to the charged Higgs is appreciable (35\%). We can see that the
branching ratios for events with large numbers of jets, such as the 
6 jets-0 lepton, 5 jets-0 lepton, and the 4 jets-1 lepton cases, have 
significant drops in their branching ratios. On the other hand, events
with fewer numbers of jets have an increase in their branching ratios as 
demonstrated by the 3 jets-0 lepton and 2 jets-0 lepton cases.

The CDF Collaboration has recently performed a search for new particles
(``$X$'')
decaying into $b \overline{b}$ produced in association with $W$ bosons
decaying into electrons or muons [\ref{CDF2}]. 
Specifically, they selected events that 
contain an electron or muon and two jets, at least one of which is
b-tagged. 
Their main motivation was to look
for W + SM Higgs events, but presumably the acceptances are roughly the
same for the W + charged Higgs production. 
We can obtain events with this signature when one top quark decay
to a $W$ which then decays leptonically and the other top quark decays
to a charged Higgs which then decays to a tau whose decay products fail
to satisfy the cuts. This would leave us with two b-jets and a 
charged lepton. The branching ratios for such events are depicted in 
Fig.~5. As the graph demonstrates, the branching ratio for this
decay mode increases dramatically as $M_{H^\pm}$ decreases and the rate
of top quark decay via the charged Higgs increases. For example, with
$M_{H^\pm} = 110$\,GeV and $\tan \beta = 65$, the branching ratio for
this mode is 6.6\%. The standard model expectation is about 3.8\%, so
we expect an excess cross section for this mode of about
0.2\,pb  assuming $\sigma_{t \overline{t}} = 7.5$\,pb.
Using the mean value of the top quark pair production
cross section reported by CDF, $\sigma_{t \overline{t}} = 7.5$\,pb, this 
means that the production cross section for this mode is about 0.5\,pb.
The CDF results set a 95\% C.L. upper limit on 
$\sigma_{WX} \cdot B(X \rightarrow b \overline{b})$ 
of 20\,pb for $M_X = 110$\,GeV. Factoring in the W
decay rate to $e$'s and $\mu$'s gives a 2 b-jets and 1 lepton cross section 
of about 5\,pb. Thus the CDF results do not impose any real restriction
on the charged Higgs decays of the top quark. 
As the total integrated
luminosity increases for runs at the upgraded Tevatron, 
the charged Higgs signal may be observable in this mode. The absence of this
signal will exclude some region of the $M_{H^\pm}$-$\tan \beta$ parameter
space.

\section{Conclusion}

If the charged Higgs boson is light enough, it can provide an additional
decay channel for the top quark. 
It can thereby potentially be detected at the Tevatron through
top quark pair production. 
The presence of this charged Higgs production
at the Tevatron would manifest itself through a change in the
branching ratios for the various final states available to top pair 
production.
Indeed, we have seen that the inclusion of the decay $t \rightarrow H^+ b$
leads to a decrease in events with large numbers of jets for a given 
number of leptons. In particular, this is true for the 2-jets dilepton mode
and the 4-jets and lepton mode; both of which are important low background
channels for investigating top quark production at the Tevatron.
There is likewise a general increase in events with smaller numbers of jets
as the rate for $t \rightarrow H^+ b$ increases. 

Current CDF data on the 2 b-jets and 1 lepton channel do not pose any real
restriction on the charged Higgs decays of the top quark. On the other hand,
data from an upgraded Tevatron could potentially detect the charged Higgs
boson in this mode or rule out some significant portion of the 
$M_{H^\pm}$-$\tan \beta$ parameter space.

\section*{Acknowledgments}

We thank  Roberto Vega for discussions. This work was supported in part
by U.S. Department of Energy grants DE-FG013-93ER40757 and 
DE-FG02-94ER40852.

\newpage

\begin{figure}
\begin{minipage}[b]{\textwidth}
\includegraphics[width=\textwidth,viewport=30 220 477 543,clip]{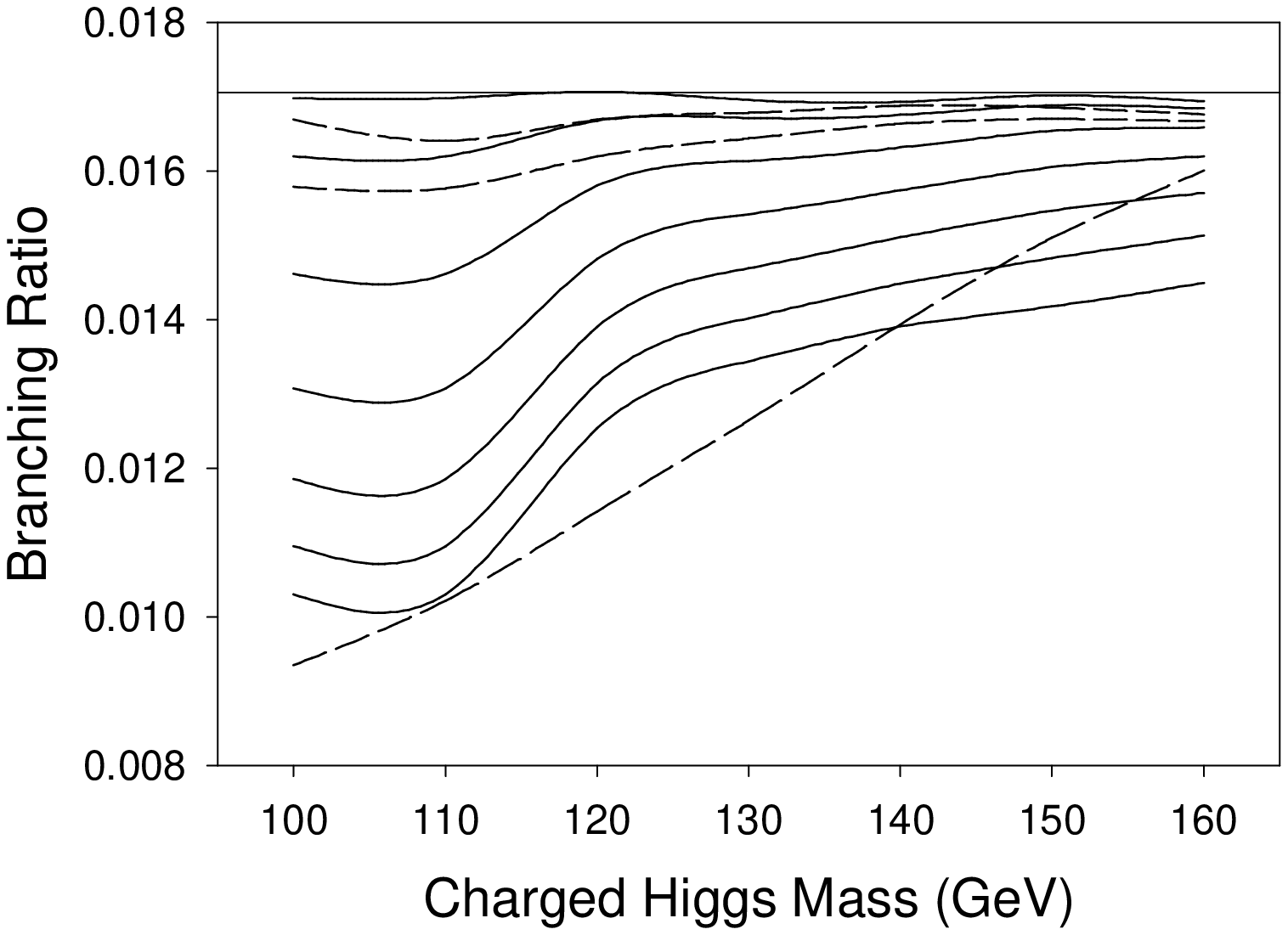}
\caption{Branching fractions as a function of the charged Higgs mass for
the 2 jets and 2 leptons case. From top to bottom, the solid curves are
for $\tan \beta =$ 5, 20, 35, 50, 65, 80, and 95, respectively. 
From top to bottom, the dashed curves are for $\tan \beta =$ 3, 1.5, and 1.
The horizontal line indicates the SM expectation.}
\end{minipage}
\end{figure}

\newpage

\begin{figure}
\centering
\begin{minipage}[b]{0.45\textwidth}
\centering
\subfigure[The 4 jets and 1 lepton case. From top to bottom, the solid
  curves are for $\tan \beta =$ 5, 20, 35, 50, 65, 80, and 95. The upper
  dashed curve is for $\tan \beta = 3$, while the lower one is for 
  $\tan \beta = 1$.]  {%
  \includegraphics[width=0.98\textwidth,viewport=40 220 477 542,clip]{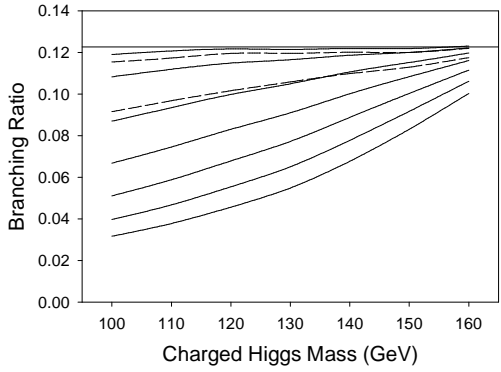} }%
\end{minipage}%
\hspace{0.1\textwidth}%
\begin{minipage}[b]{0.45\textwidth}
\centering
\subfigure[The 3 jets and 1 lepton case. From top to bottom, the solid
  curves are for $\tan \beta =$ 95, 80, 65, 50, 35, 20, and 5. The upper 
  dashed curve is for $\tan \beta = 1.25$, while the lower one is for $\tan
  \beta = 1$.]  {%
  \includegraphics[width=0.98\textwidth,viewport=40 240 486 562,clip]{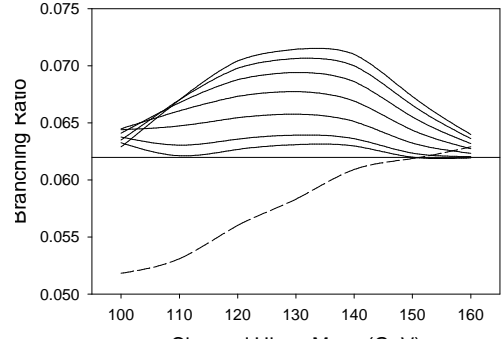}  }%
\end{minipage} \\
\begin{minipage}[b]{0.45\textwidth}
\centering
\subfigure[The 2 jets and 1 lepton case. From top to bottom, the solid
  curves are for $\tan \beta =$ 95, 80, 65, 50, 35, 20, and 5. The upper
  dashed curve is for $\tan \beta = 3$ and the lower dashed curve is for
  $\tan \beta = 1$.]  {%
  \includegraphics[width=0.98\textwidth,viewport=40 220 477 542,clip]{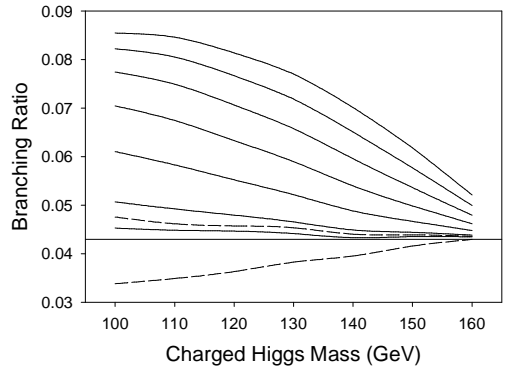}  }%
\end{minipage}
\caption{Branching fractions as a function of the charged Higgs mass
for the single lepton modes. The horizontal line in each plot shows the
SM expectation.}
\end{figure}

\newpage

\begin{figure}
\centering
\begin{minipage}[b]{0.45\textwidth}
\centering
\subfigure[The 6 jets and 0 lepton case. From top to bottom, the solid curves 
  are for $\tan \beta =$ 5, 20, 35, 50, 65, 80, and 95. The upper dashed
  curve is for $\tan \beta = 3$ and the lower dashed curve is for
  $\tan \beta = 1$.]  {%
  \includegraphics[width=0.98\textwidth,viewport=40 220 477 542,clip]{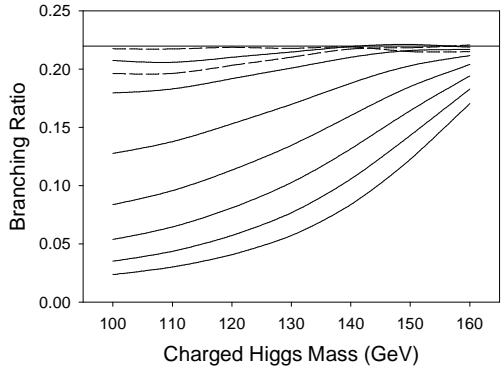}  }%
\end{minipage}%
\hspace{0.1\textwidth}%
\begin{minipage}[b]{0.45\textwidth}
\centering
\subfigure[The 5 jets and 0 lepton case. From top to bottom, the solid curves
  are for $\tan \beta =$ 5, 20, 35, 50, 65, 80, and 95.]  {%
  \includegraphics[width=0.98\textwidth,viewport=40 220 477 542,clip]{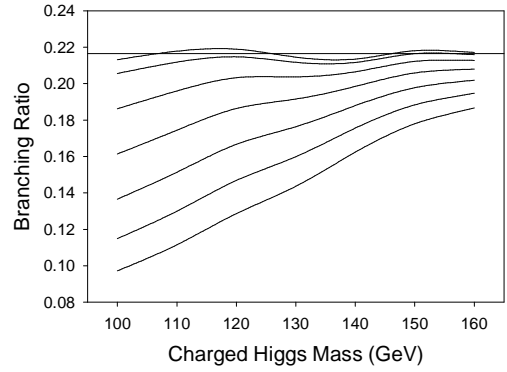}  }%
\end{minipage} \\
\begin{minipage}[b]{0.45\textwidth}
\centering
\subfigure[The 4 jets and 0 lepton case. From top to bottom, the solid curves
  are for $\tan \beta = $ 95, 80, 65, 50, 35, 20, and 5. The dashed curves are
  for $\tan \beta = 1$ and $\tan \beta = 3$.]  {%
  \includegraphics[width=0.98\textwidth,viewport=40 220 477 542,clip]{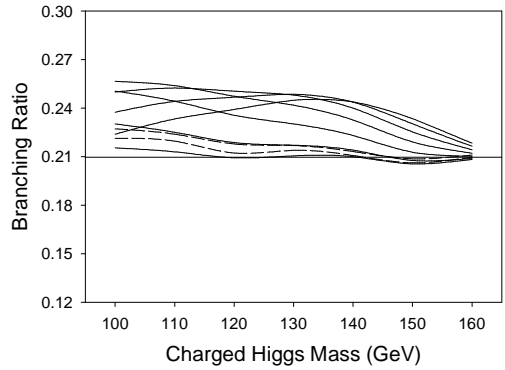}  }%
\end{minipage}
\caption{Branching fractions as a function of the charged Higgs mass
for the purely hadronic modes. The horizontal line in each plot shows the
SM expectation.}
\end{figure}

\addtocounter{figure}{-1}
\begin{figure}
\addtocounter{subfigure}{3}
\centering
\begin{minipage}[b]{0.45\textwidth}
\centering
\subfigure[The 3 jets and 0 lepton case. From top to bottom, the solid curves
  are for $\tan \beta = $ 95, 80, 65, 50, 35, 20, and 5. The upper dashed
  curve is for $\tan \beta = 1$ and the lower one is for 
  $\tan \beta = 3$.]  {%
  \includegraphics[width=0.98\textwidth,viewport=40 220 477 542,clip]{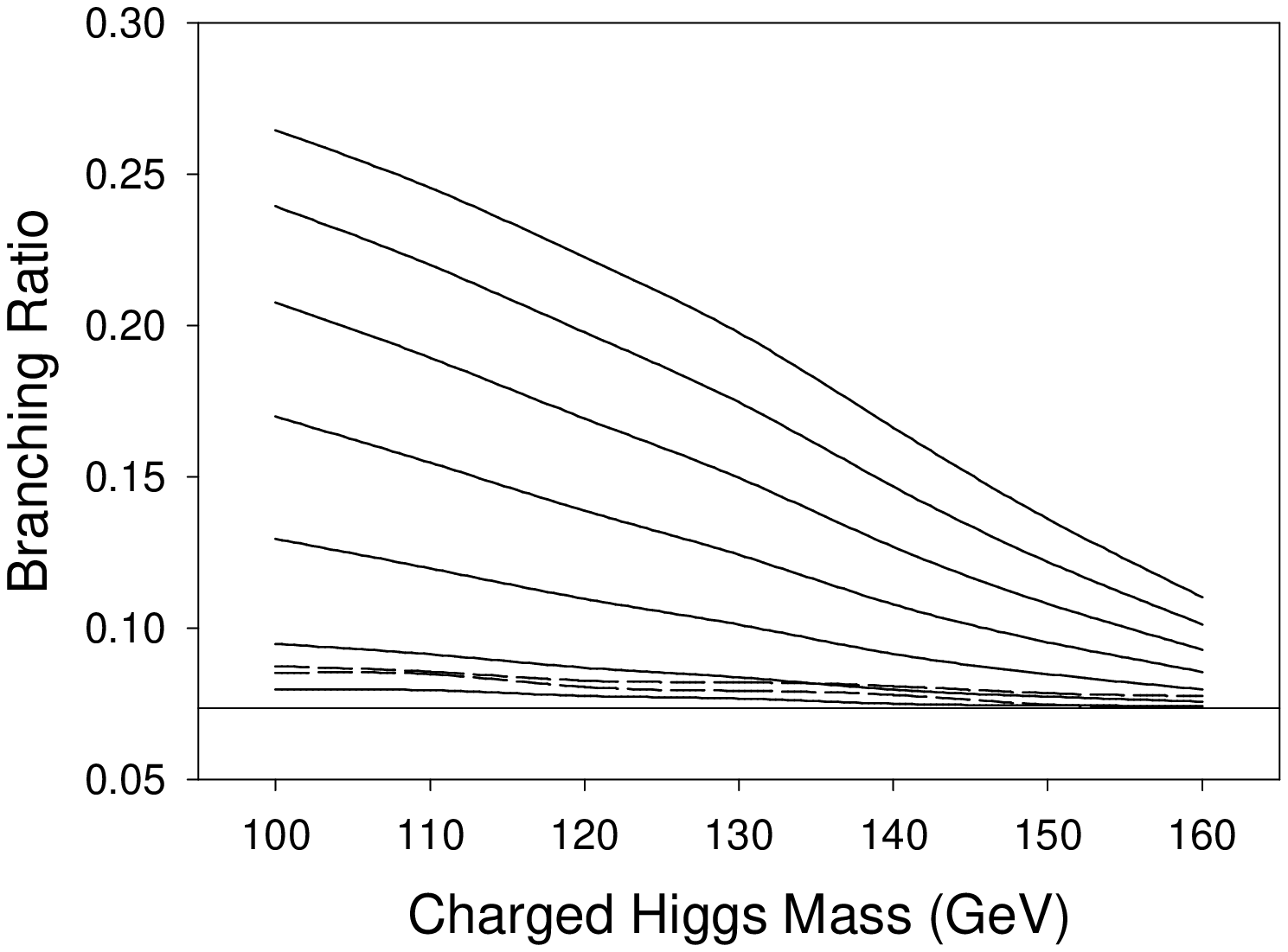}  }%
\end{minipage}%
\hspace{0.1\textwidth}%
\begin{minipage}[b]{0.45\textwidth}
\centering
\subfigure[The 2 jets and 0 lepton case. From top to bottom, the solid curves
  are for $\tan \beta = $ 95, 80, 65, 50, 35, 20, and 5. The upper dashed
  curve is for $\tan \beta = 1$ and the lower one is for 
  $\tan \beta = 3$.]  {%
  \includegraphics[width=0.98\textwidth,viewport=40 220 477 542,clip]{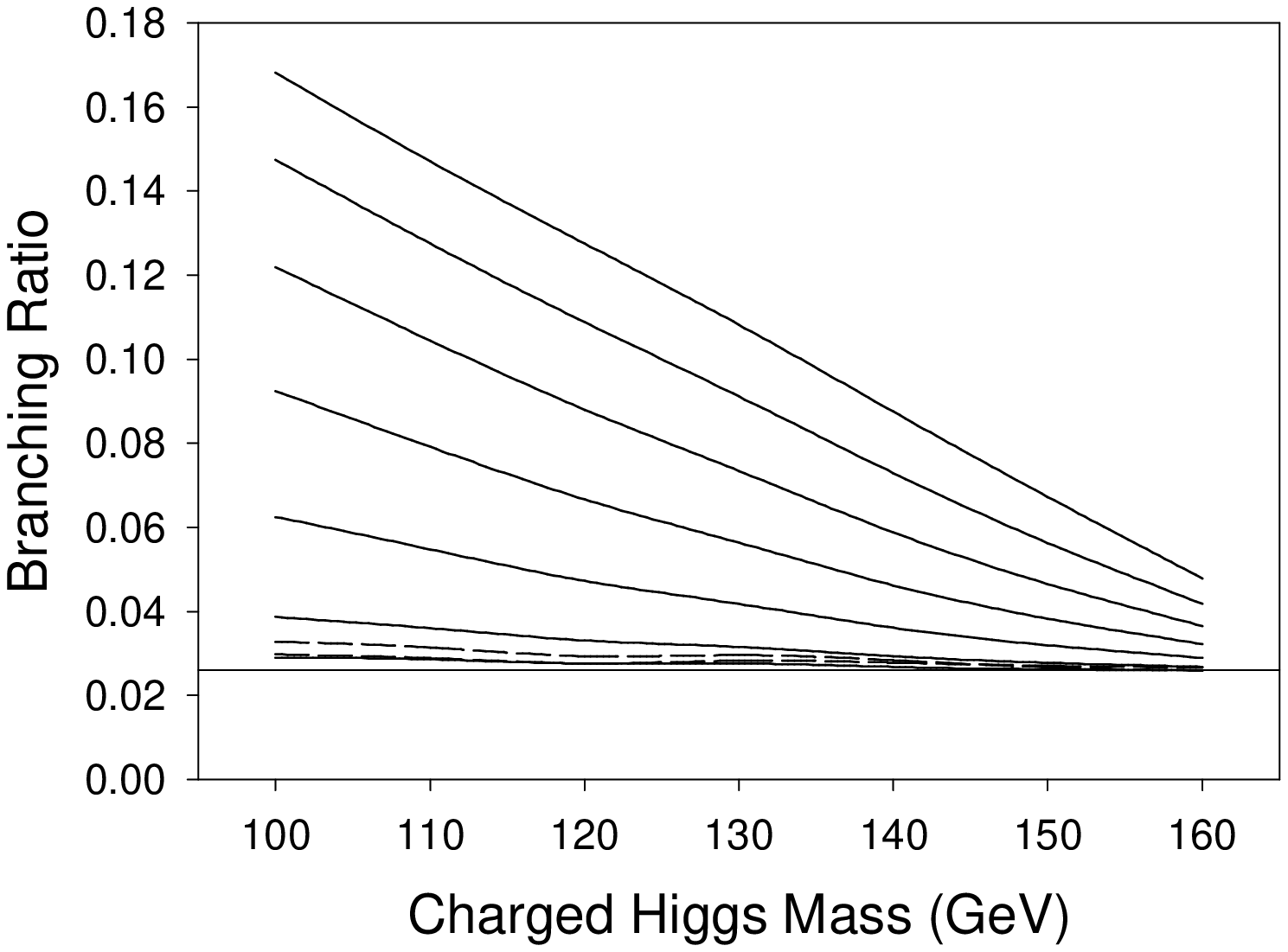}  }%
\end{minipage}
\caption{continued.}
\end{figure}

\newpage

\begin{figure}
\centering
\begin{minipage}[b]{0.45\textwidth}
\centering
\subfigure[The standard model expectation.]  {%
  \includegraphics[width=0.98\textwidth,viewport=75 230 456 548,clip]{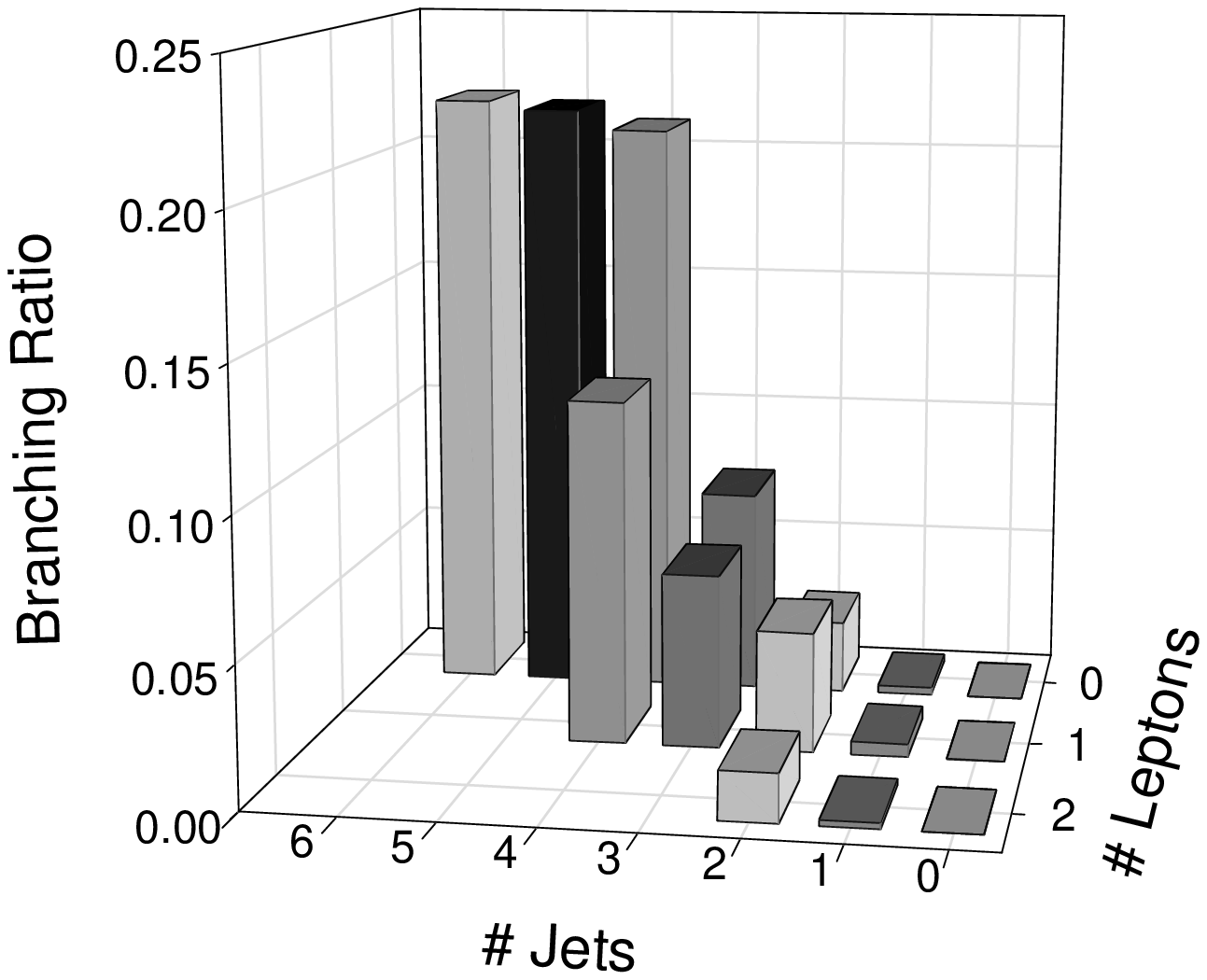}  }%
\end{minipage}%
\hspace{0.1\textwidth}%
\begin{minipage}[b]{0.45\textwidth}
\centering
\subfigure[The MSSM case where $\tan \beta = 3$ and $M_{H^\pm} = 110$\,GeV.]
{%
  \includegraphics[width=0.98\textwidth,viewport=75 230 456 548,clip]{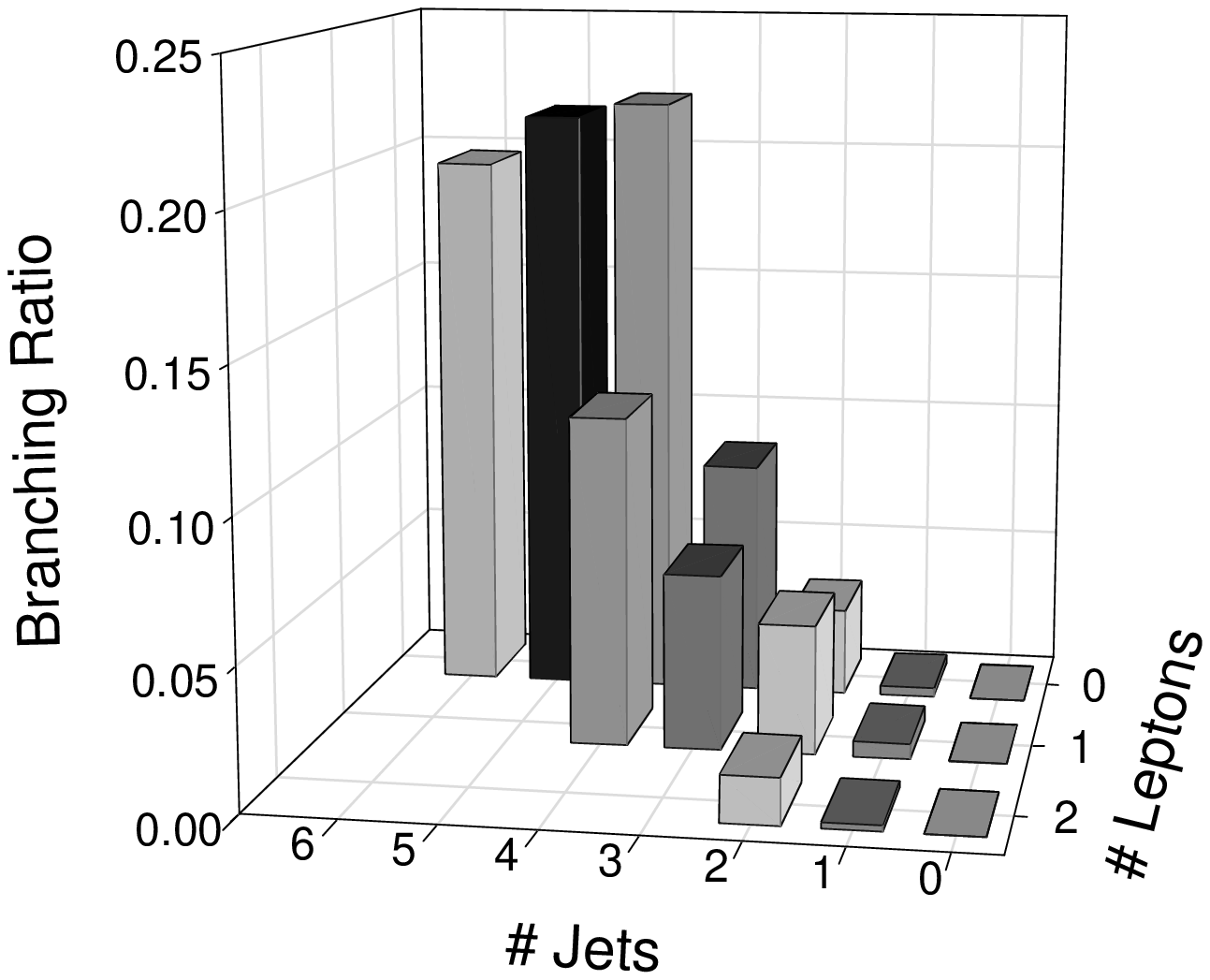}  
}%
\end{minipage} \\
\begin{minipage}[b]{0.45\textwidth}
\centering
\subfigure[The MSSM case where $\tan \beta = 50$ and $M_{H^\pm} = 110$\,GeV.]
{%
  \includegraphics[width=0.98\textwidth,viewport=75 230 456 548,clip]{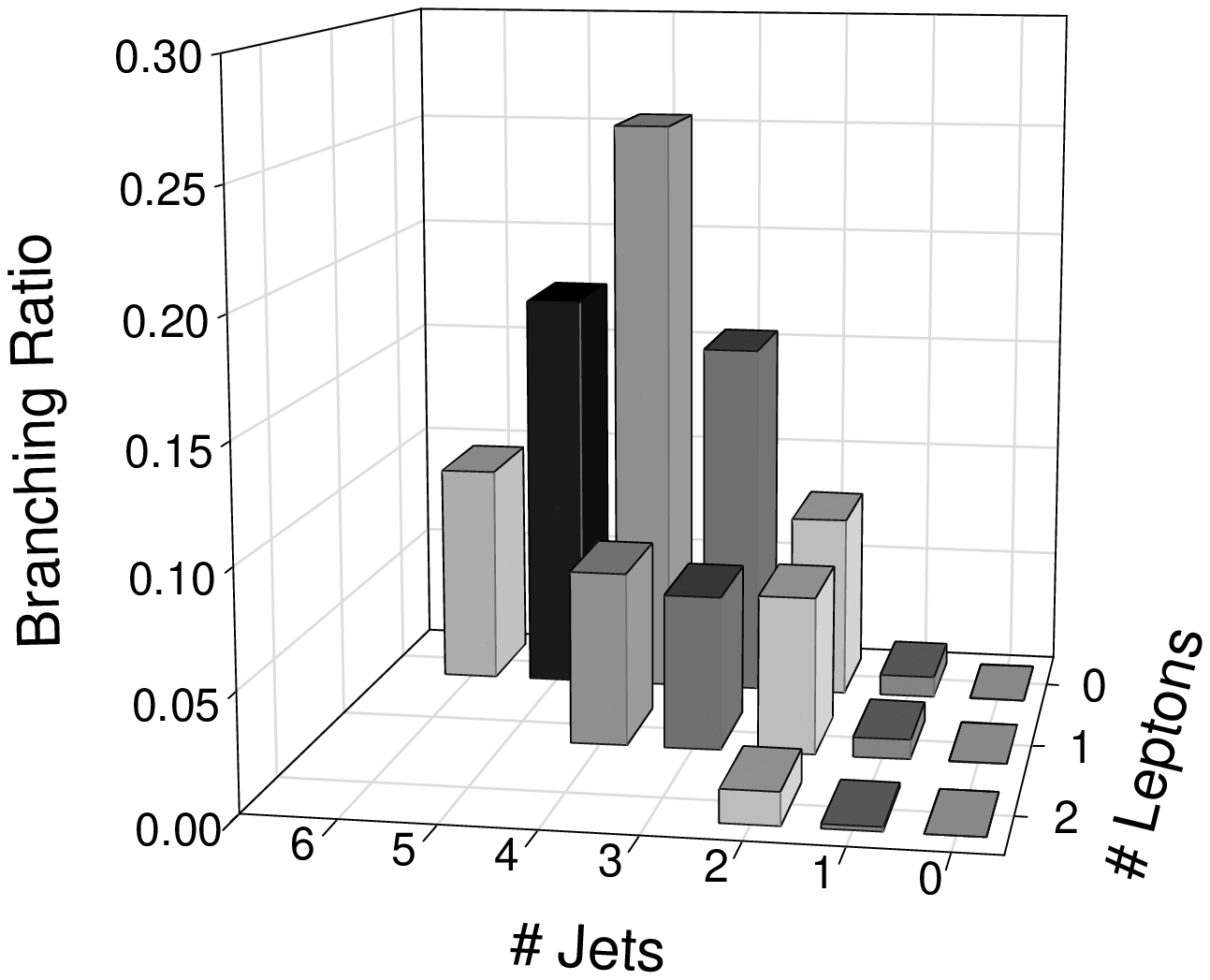}
}%
\end{minipage}
\caption{Histograms of the branching ratios for the possible decay
modes in top quark pair production.}
\end{figure}

\newpage

\begin{figure}
\begin{minipage}[b]{\textwidth}
\includegraphics[width=\textwidth,viewport=50 120 490 448,clip]{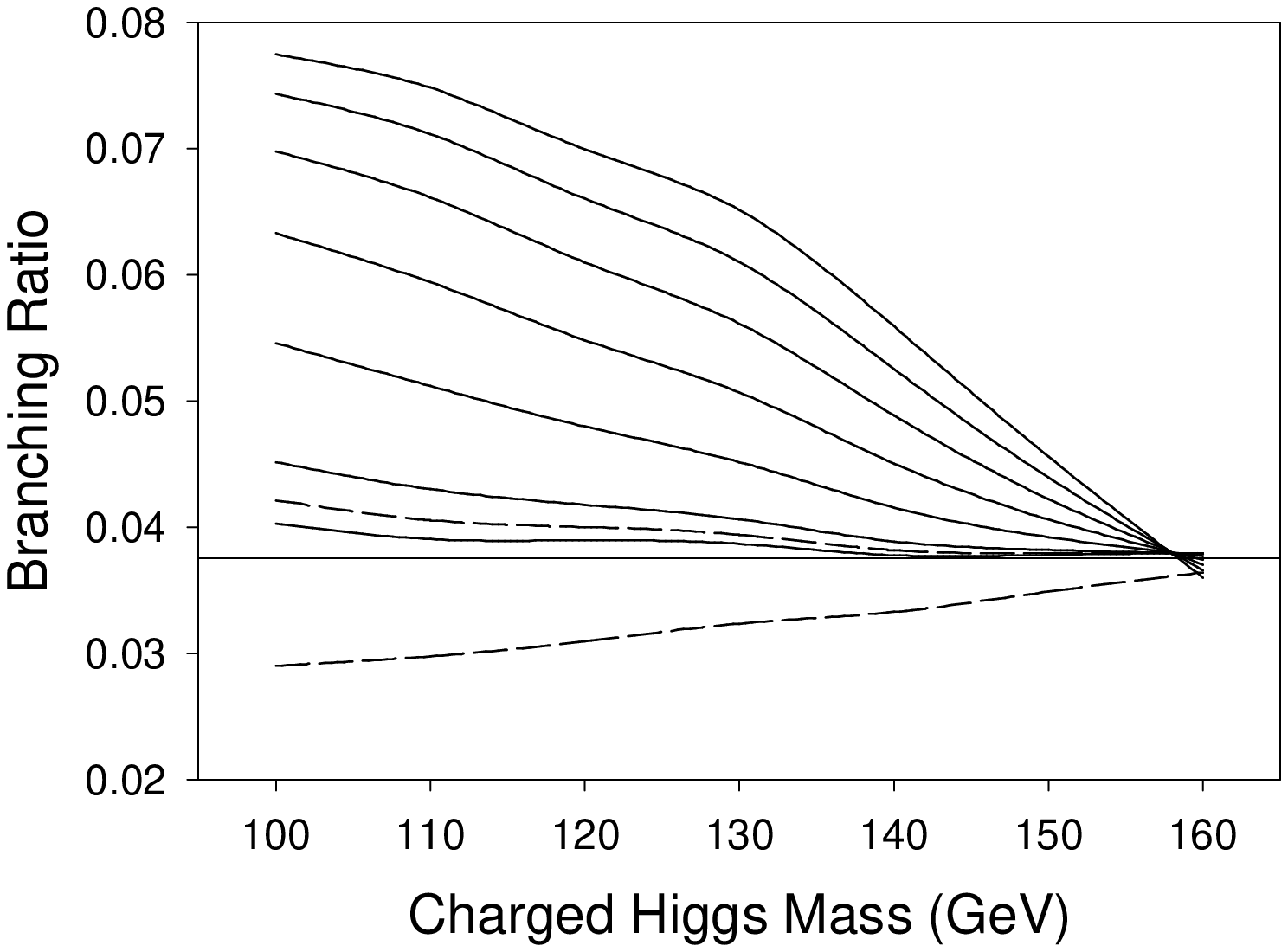}
\caption{Branching fractions as a function of the charged Higgs mass
for the two b-tagged jets and single lepton mode. From top to bottom, the
curves are for $\tan \beta =$ 95, 80, 65, 50, 35, 20, and 5. The dashed
lines are for $\tan \beta =$ 3 and 1. The solid horizontal line depicts
the SM expectation.}
\end{minipage}
\end{figure}

\end{document}